# Study of amplified emission in polycrystalline ZnO below characteristic temperature

Sanjiv Kumar Tiwari [1,2]

[1] *Department of Physics, Indian Institute of Technology Kanpur, Kanpur- 208016, India*

[2] *Department of physics of complex systems, Weizmann Institute of Science, Rehovot, 76100 Israel*

## ABSTRACT

We report on amplified emission in polycrystalline ZnO below 100K. At 6K emission is due to free exciton $FX^{n=1}_A$ (3.378 eV), bound exciton $D_1X_A$ (3.347 eV), donor acceptor pair DAP (3.275 eV) and longitudinal phonon replica of free excitons ($FX^{n=1}_A$ -mLO, m= 1,2,3) respectively. Peak intensity of $D_1X_A$ and $FX^{n=1}_A$ -1LO transition increases non-linearly with increase of excitation intensity due to exciton-exciton scattering. Peak position of $D_1X_A$ shows red shift while $FX^{n=1}_A$ -1LO shows blue shift with increase of excitation intensity. Fraction of exciton taking part in emission process and radiative lifetime of exciton decreases with decrease of temperature .Threshold value of excitation for $D_1X_A$ decreases exponentially with decrease of temperature upto 100K. Whereas, below 100K no functional behaviour is observed. Threshold for $FX^{n=1}_A$-1LO emission was observed up to 75K, at higher temperature it is mixed with $D_1X_A$. Time delayed photoluminescence measurement reveals that amplified emission decay faster than spontaneous emission.

---

[1] *correspondence should be address at: tiwarisanjiv@gmail.com*

**Introduction.**

ZnO is a wide band gap material. It has band gap energy of 3.37 eV and exciton binding energy of 60 meV at room temperature. Because of large binding energy of excitons it is expected that ZnO can sustain excitonic emission at higher temperature. ZnO nano-sheets can sustain excitonic emission till 870K [1]. Emission process in ZnO is band to band transition and direct recombination of exciton can not be used to establish amplified/stimulated emission. Because system of an exciton coupled to photon is equivalent to two coupled oscillator thus, participation of third field is necessary to get amplified/stimulated emission, such as bi-exciton decay, exciton-exciton scattering (ex-ex), exciton-electron scattering (ex-el) and/or formation of electron hole plasma (EHP) [2,3,4]. Bagnall observed distinct free electron peak and EHP peak with increase of excitation intensity in temperature range of 300-550K [5]. Room temperature stimulated emission in ZnO and other materials have been observed in many cases [6, 7, 8]. However, investigation on the temperature dependence of stimulated emission from ZnO reveals the characteristic temperature ($T_c$) which reflect the quality of high performance of the laser. $T_c$ is found to be 90K and 127K in temperature range of 294K -377K and 300-570K respectively [9, 10]. Almost in all the case of ZnO samples threshold intensity for non-linear emission decreases exponentially with decrease of temperature and $T_c$ is in the range of 90-130K. In present article, we start with intrinsic process in the intermediate density regime and assume weak exciton-photon interaction e.g we neglect the polariton concept through out the paper to discuss the amplified emission process in ZnO below characteristic temperature. This article is devoted to amplified emission in polycrystalline ZnO due to $D_1X_A$ and $FX^{n=1}_A - 1LO$ transition below $T_c$, fraction of exciton taking part in emission process and comparison of radiative life time below and above threshold together with nonlinear increase and blue shift of $FX^{n=1}_A - 1$ LO peak with increase of excitation intensity.

**Experimental setup and method.**

ZnO sample were prepared by cold press of ZnO powder 99.99% (Sigma-Aldrich) at pressure of 6 Ton and sintered at 1000 $^0$C in air for five hours. Steady-state photoluminescence was carried out with the sample placed in a closed cycled cryostat in the temperature range of 100-6 K. Third harmonic of Nd:YAG ( DCR-4G, Spectra Physics λ=355 nm, 4 nano-second pulse duration, 10 Hz repetition rate, 2 mJ/pulse )  was used as excitation source. Excitation area was circle of diameter 400 micrometer. The emission light was collected by a lens, imaged onto fiber-coupled monochromator and detected by Intensified Charge Couple Device (ICCD DH720, Andor Technology). ICCD was operated in gating mode and emission was collected for 1 microsecond after laser pulse. The x-ray diffraction ( XRD) spectra of ZnO sample were taken at room temperature using CuK$_\alpha$ source (λ=1.54 A$^0$) to get the structural properties of the sample. XRD spectra of ZnO show that sample is polycrystalline with average grain size of 28 nm. The excited exciton density $n_{ex,}$ within focal volume by UV illumination was calculated using rate equation $n_{ex} = \frac{\eta \lambda}{dhc} \tau_{Fx} p$ . Where *P* is the excitation intensity, *λ* is the excitation wavelength, *h* is plank's constant, *c* is the velocity of light in vacuum, *d* ( ≈ 500 nm) is the thickness of active layer, τ$_{Fx}$ (≈ 1x 10$^{-9}$ sec ) is radiative life time of free excitons and *η (*=0.40) is coupling efficiency [11,12]. This leads to the exciton density of 10$^{18}$-10$^{20}$ cm$^{-3}$ within focal volume with increase of excitation intensity from 9.10 kW-cm$^{-2}$ to 3.87 MW-cm$^{-2}$ respectively. This calculation is in agreement with Klingshrin, who propose that maximum generation rate of exciton with excitation power of 500 kW-cm$^{-2}$ in fundamental edge with UV illumination can be taken as 10$^{29}$ cm$^{2}$-sec$^{-1}$. This assumption also gives exciton density of order of 10$^{20}$ cm$^{-3}$ [13, 14, 15] . Further assuming excitation area as cylinder of length α$^{-1}$ (α=2 x 10$^5$ cm$^{-1}$ at 355 nm) and diameter 400 micron e.g excitation volume ≈3 x 10$^{-9}$ cm$^3$ and exciton density ≈ 10$^{20}$ cm$^{-3}$ gives average separation between excitons (<r>) of about 100 A$^0$ . This implies that <r> is comparable of exciton diameter (35 A$^0$). These calculations suggest that exciton-exciton interaction is strong in our case and formation of new complex like bi-exciton or exciton molecules is less likely.

Whereas, formation of electron–hole plasma (EHP) is negligible because it requires exciton density of order of $10^{23}$ cm$^{-3}$.

**Results and discussion.-** Figure 1 shows the emission/Photoluminescence(PL) profile of polycrystalline ZnO at 6K with increase of excitation intensity from 9.1 kW-cm$^{-2}$ to 3.87 MW-cm$^{-2}$. Emission profile at 6K reveals free excitonic peak FX$^{n=1}_A$ (3.378 eV), FX$^{n=2}_A$ (3.414 eV), FX$^{n=1}_B$ (3.391 eV) together with first, second and third longitudinal phonon ( FX$^{n=1}_A$ -mLO, m= 1, 2,3) replica of FX$^{n=1}_A$ at 3.306 eV, 3.236 eV and 3.167 eV respectively, inset (a). Peak positions of free exciton are in agreement with earlier reported values [16, 17]. Bound exciton peak D$_1$X$_A$, 1$^{st}$ phonon replica of bound exciton ( D$_1$X$_A$ -1LO), donor acceptor pair (DAP) transition and free electron acceptor (e, A) peak occurs at 3.347 eV, 3.275 eV, 3.206 eV and 3.187 eV respectively. The details about origin and position of all peaks can be found some where else [16]. Excitation intensity was varied with help of linear polarizer by varying the polarizer angle, inset (b) while, inset (c) shows the collection geometry of emission. It is evident from figure 1 that with increase of excitation intensity all lower energy peaks get merged together and remaining dominating lines are D$_1$X$_A$ and FX$^{n=1}_A$ -1 LO . Whereas, at 200K D$_1$X$_A$ and FX$^{n=1}_A$ -1LO also get merged into single broad peak (figure not shown). Figure 2 shows the variation of PL peak intensity with excitation intensity in temperature range of 6-100K. PL peak intensity increases non-linearly with increase of excitation intensity and there exist threshold for nonlinearity. Threshold value is shown as dotted arrow. The position of PL peak is given as [18, 19],

$$\hbar\omega_{PL} = E_{FX} - E_b^{FX}(1-\frac{1}{n^2}) - \frac{3}{2}kT, n>1 \ldots \qquad 1$$

Where, $E_{FX}$ is energy of free exciton, $E_b^{FX}$ is binding energy of free exciton and 3/2 kT is kinetic energy term. Further taking into account that below 100K bound exciton dominates hence, equation 1 can be re-written as [20, 21],

$$\hbar\omega_{PL} = E_{D_1X_A} - E_b^{D_1X_A}(1-\frac{1}{n^2}), n>1 .. \qquad 2$$

Here, $E_{D_1X_A}$ is energy of bound exciton, $E_b^{D_1X_A}$ is binding energy of exciton with neutral donor, and kinetic energy term can be neglected for bound exciton. Taking $E_{D_1X_A}$ =3.347 eV, $<E_b^{D_1X_A x}>$=16 meV Red shift of $E_{D_1X_A}$ is 5 meV, which is in agreement with experimental value 6-8 meV in temperature range of 6-100K . Following [15], it has been shown that exciton-exciton interaction line known as **P** band luminescence increases super linearly with increase of excitation intensity and particularly EHP line is enhanced due to stimulation effect, which also produces line narrowing. Surprisingly, we did not observe narrowing in line width. Although, non-linearity in emission is clearly seen in figure 2. Spectral width of emission gets broader with increase of exciton intensity and finally settled at constant value. This is due to strong coupling of D $_1$X$_A$ with FX$^{n=1}{}_A$ .The longitudinal separation between A and B exciton is 13 meV almost equal to the line width of D $_1$X$_A$. In addition, since scattering of A and B exciton is equally probable therefore, it is expected that A and B exciton series get mixed together during exciton-exciton scattering process. As a result, emission line gets broader with increase of excitation intensity. Although, exciton-electron interaction is also possible at low temperature and higher excitation intensity. In order to show that exciton-exciton interaction is dominating one, figure 3 shows the shift of D $_1$X$_A$ line (ΔE) with increase of excitation intensity at different temperature. It shows power law behavior as $\Delta E \approx P^{0.87\pm.03}$ at 6K, whereas for exciton-electron interaction process power law is $\Delta E \approx P^{2/3}$ [14]. Further, discussion on amplified/stimulated emission is incomplete without pointing out the radiative lifetime. It is worth mentioning that not all excited exciton recombine radiatively during emission process, only exciton close to the Brillouin-zone centre recombines radiatively . Figure 4(a) shows the fraction of exciton *r(T)* within the spectral width *ΔI$_{PL}$* that contribute to radiative recombination. *r(T)* was calculated assuming a parabolic exciton band and Maxwell-Boltzmann distribution [21].

$$r(T) = \frac{2}{\sqrt{\pi}} \int_0^{\Delta I_{PL}(T)/k_B T} \sqrt{\varepsilon} e^{-\varepsilon} d\varepsilon \qquad 3$$

Where, $k_B$ is Boltzmann constant, and $\Delta I_{PL}$ is FWHM of D$_1$X$_A$ peak. Theoretically, it has been shown that radiative lifetime ($\tau$) of exciton is related to the width of its emission spectrum as

$$\tau \propto (E_B)^{-1}(M/\mu)\Delta I_{PL}(T)/r(T) \qquad 4$$

Where, $E_B$ is exciton binding energy, $M$ and $\mu$ are the total mass and reduced mass of exciton respectively. Combining equation 3 and 4

$$\tau = \gamma \Delta I_{PL}(T)/r(T) \qquad 5$$

Where, γ is temperature independent constant. Figure 4 (b) shows the variation of $\Delta I_{PL}/r(T)$ versus excitation intensity. $r(T)$ is almost constant with excitation intensity at 6K and 25K then it decreases with increase of temperature. This indicates homogeneous distribution of thermal/kinetic energy of excitons within their energy state. Whereas, below 1MW-cm$^{-2}$ of excitation intensity and above 50K, $r(T)$ first decreases then increase with further increase of excitation. This behavior could be due to flow or redistribution of carrier into different excitonic states. It is also evident from figure 4(b) that radiative life time also depends on excitation intensity and it increases with increase of temperature for a fixed excitation intensity. This is in agreement with earlier reported time resolve experiments [18, 21]. Increase of $r(T)$ or radiative life time is directly related to the oscillator strength. and radiative decay time is inversely proportional to the oscillator strength, which is reduced when kinetic energy range of effective exciton recombination is broadened. Hence, more power is needed to get required level of emission at higher temperature. This is also the obvious reason for increase of threshold with increase of temperature. Figure 4(c) shows the variation of threshold intensity with sample temperature. For free exciton emission (temperature above 100K) the threshold decreases exponentially with decrease of sample temperature. However, below 100K it decays fast. The decrease of threshold for 5K≤T≤100K can be explained by the fact that at low temperature, the bound excitons are an effective recombination channel and oscillator strength of bound exciton is much higher than free excitons.

In order to study the decay time of excitons, we used ICCD in gating mode which has time resolution of 3 nano-second. As stated in literature, the decay time of excitonic luminescence is of the order of 1 nano-second or even less (order of pico-seconds) and our experimental setup has insufficient time resolution for correct decay measurement. However, a comparison of decay process below and above threshold can be made. For comparison of decay time below threshold and above threshold of excitation intensity, emission was recorded for 3 nano-seconds in step of 3 nano-seconds time delay with respect to laser pulse until emission vanishes.. Although, exciton decay is bi-exponential [18] with decrease of temperature but due to low resolution of ICCD we consider only single exponential decay and decay time was estimated using $I(t) = I(0)\exp(-t/\tau)$ . It is evident from figure 5 that decay time decreases from 3 to 2 nanosecond with increase of excitation intensity from 0.14 to 3.87 MW-cm$^{-2}$ . This is in agreement with earlier time resolve measurement by Teke *et-al* [16]. They observe decay of exciton lifetime from 3 to 2.5 nanosecond with increase of excitation intensity from 54 µJ-cm$^{-2}$ to 540 µJ-cm$^{-2}$. Although for donor bound exciton life was almost constant. In our case it also agrees well at 6K which has life time of 1.85 and 2 nano seconds. In addition to bound exciton recombination one also detect the stimulated emission at low temperature for a process in which a polariton is scattered from the exciton like part to the photon like part of the dispersion spectrum by stimulated emission of longitudinal optical phonon. Figure 6 shows the variation of FX$^{n=1}_A$-1LO emission peak intensity with excitation intensity; inset shows the variation of threshold with temperature. Peak intensity of FX$^{n=1}_A$-1LO increases nonlinearly with increase of excitation intensity. This process was observed up to 75K. After 75K, it is expected that thermally activated phonon dominate the process. Threshold for FX$^{n=1}_A$-1LO process also decreases with decrease of temperature, inset of figure 6(a). Interestingly PL peak position of FX$^{n=1}_A$-1LO get blue shifted with increase of excitation intensity, figure 6(b). This is due to dependence of spectral line shape of $FX^{n=1}_A$-1LO on temperature and free exciton energy. The spectral line shape *I(LO)* of the first LO phonon replica of the free exciton has the form $I(FX-1LO) \approx E^{1/2} \exp(-\frac{E}{k_B T}) P(E)$ here *E* equals $\hbar\omega - (E_{FX^{n=1}_A} - \hbar\omega_{LO})$ and *P(E)* is the transition probability which varies as $E^m$ for the corresponding phonon assisted transition, where m is constant. At low temperatures (< 75K) *P(E)* is typically assumed to be proportional to E [22]. Therefore, the peak position of 1LO band will be shifted by $3/2 k_B T$ to the higher energy side.

**Conclusions.-** To conclude, two interesting observations (i) amplified emission of $D_1 X_A$ and $FX^{n=1}_A$-1LO, (ii) comparisons decay time below and above threshold are presented in this paper. First due to exciton-exciton scattering and increase of oscillator strength. Although it was very difficult to distinguish between exciton-exciton and exciton-electron scattering process. But thanks to power law behaviors, which helps to distinguish between two. Even, without fast devices like streak camera we were able to compare the decay time between spontaneous emission and stimulated emission and radiative lifetime as function of temperature. Second the amplified emission of $FX^{n=1}_A$-1LO is presented in terms of occupation number of phonons. Decay of threshold for $D_1 X_A$ and FX-1LO show different behavior in temperature range of 6K-100K. This needs further deep investigation on decay kinetics of $D_1 X_A$ and FX-1LO separately. Although, the radiative life time of free exciton and bound exciton are same in temperature range of 6K-100K.

**Figure Captions**

**Figure 1**. Variation of PL profile at 6K with increase of excitation intensity from 9.1 kW-cm$^{-2}$ to 3.87 MW-cm$^{-2}$. Inset (a) PL profile on log scale, inset (b) variation of excitation intensity with polarizer angle, solid line is linear part of intensity variation; inset (c) shows the collection geometry, sample (S), emission (PL), and incident laser (In), reflected laser (Ref)

**Figure 2**. Variation of peak intensity of $D_1X_A$ with excitation intensity in temperature range of 6K-100K. Solid line is linear fit to experimental data points, dotted arrow show threshold value of excitation intensity.

**Figure 3.** Variation of peak energy shift with excitation intensity. Inset shows variation of peak position with excitation intensity at different temperature.

**Figure 4.** Variation of fraction of exciton taking part in radiative process Vs excitation intensity at different temperature (a), behavior of radiative life time of $D_1X_A$ as function of excitation intensity (b), variation of threshold value with temperature. Solid line shows exponential fit to experimental data points (c)

**Figure 5**. Variation of radiative lifetime of $D_1X_A$ below threshold intensity and above threshold intensity with temperature.

**Figure 6**. Variation of peak intensity of $FX^{n=1}_A$ -1LO with excitation intensity at different temperature, inset shows variation of threshold with sample temperature a). Variation of peak position of $FX^{n=1}_A$ -1LO with excitation intensity (b).

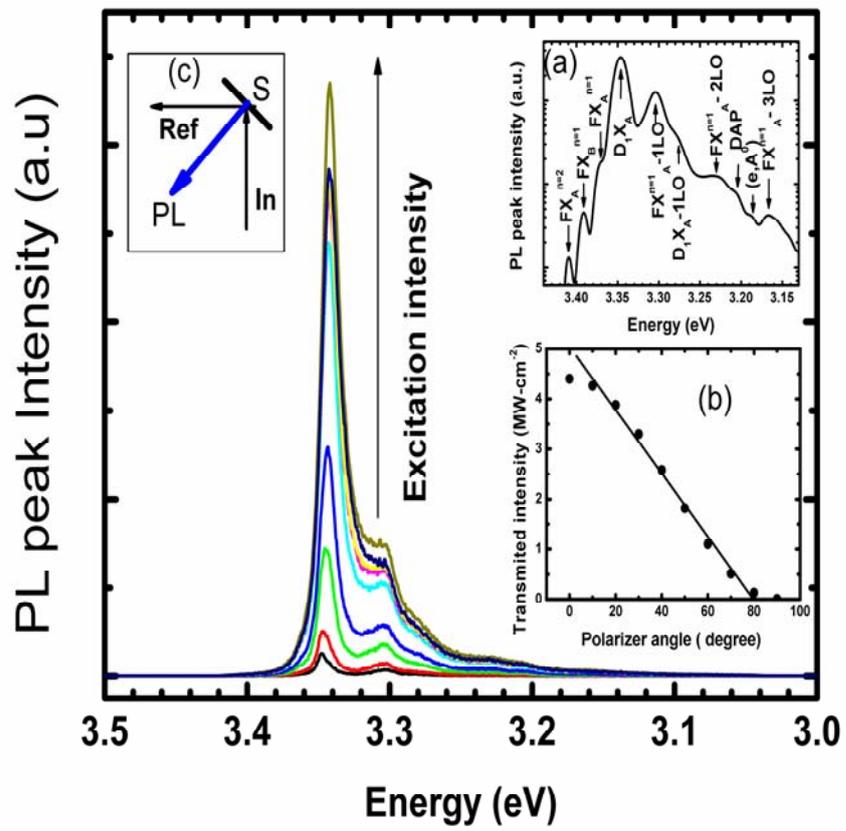

**Figure-1**

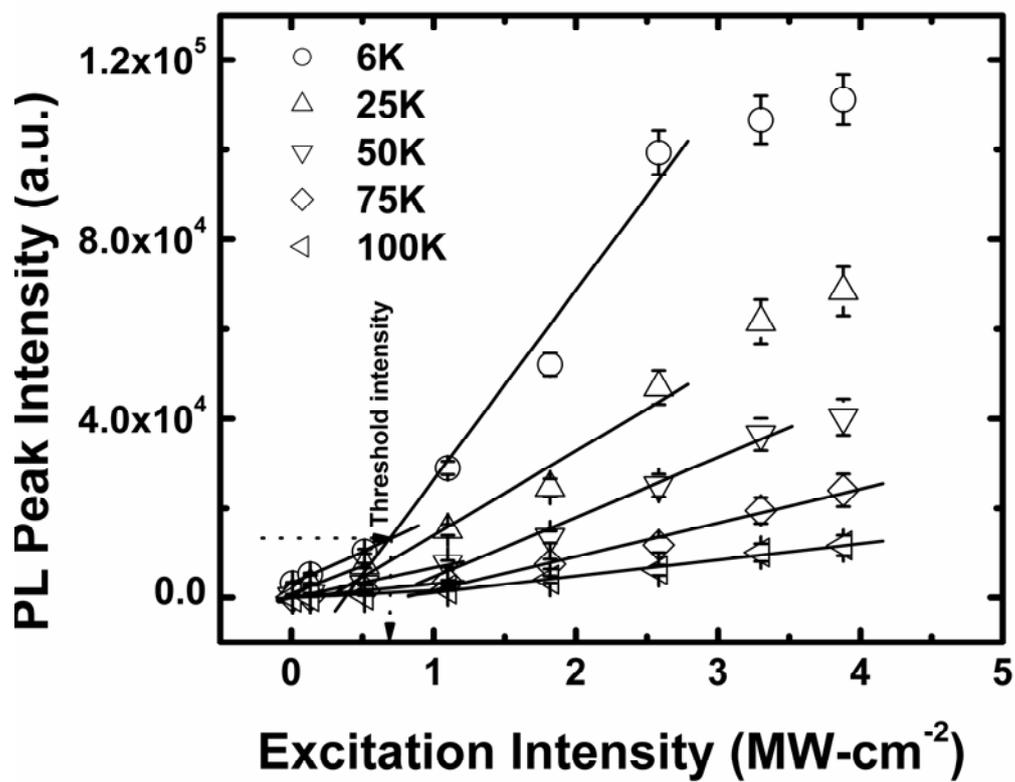

Figure 2

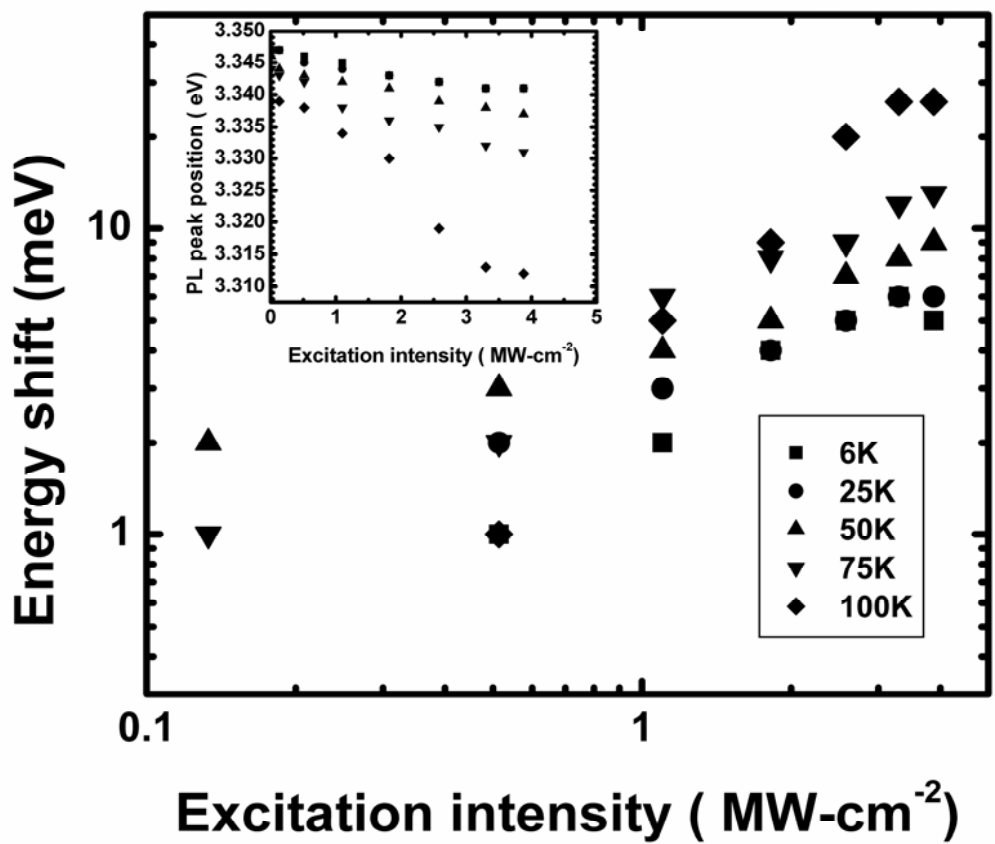

**Figure 3**

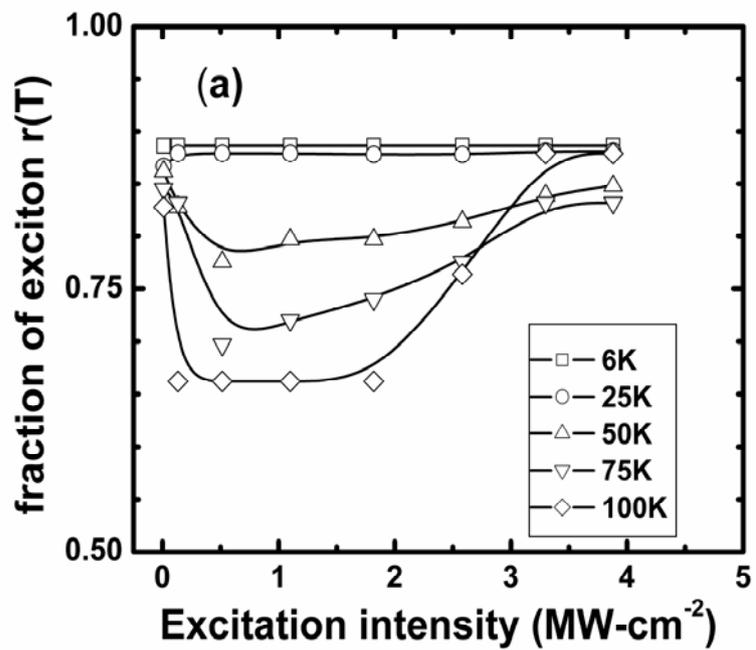

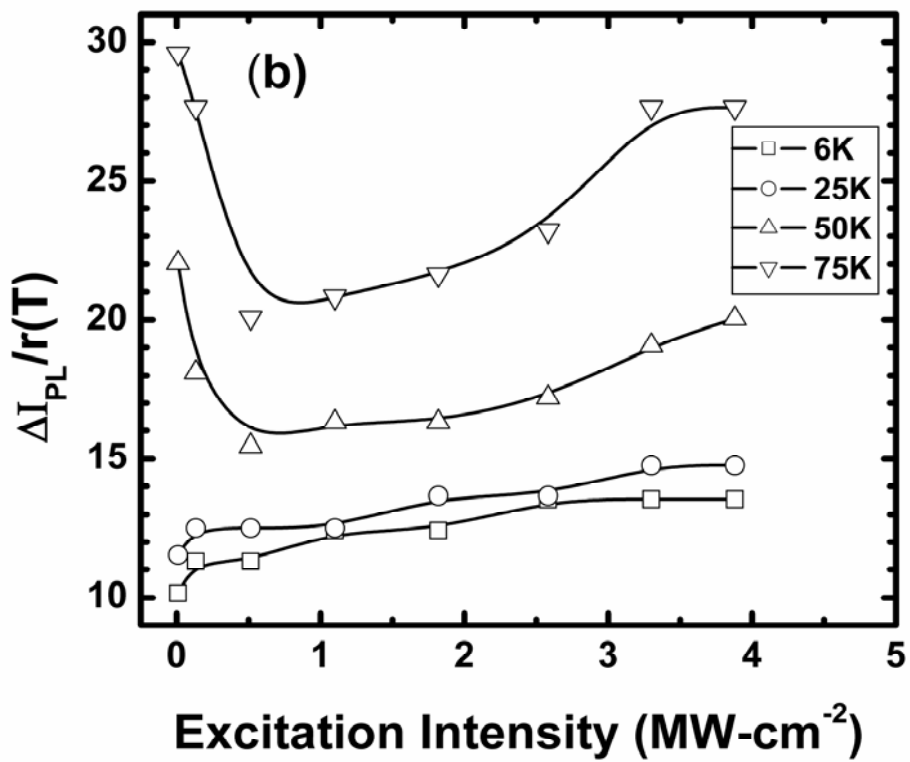

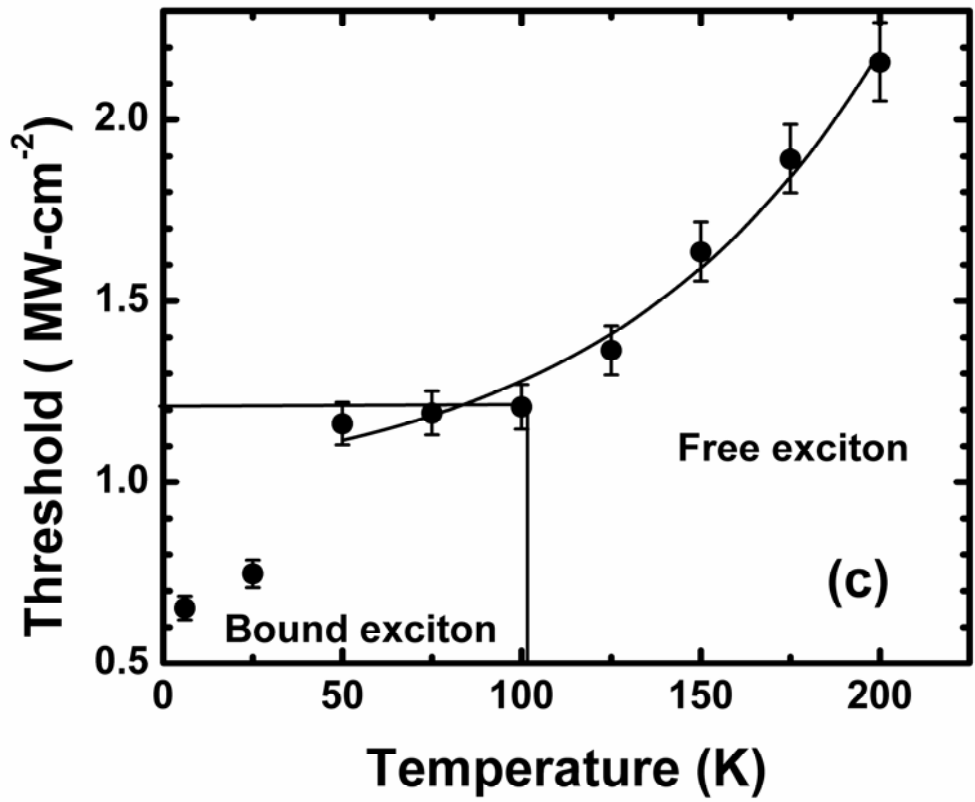

**Figure 4**

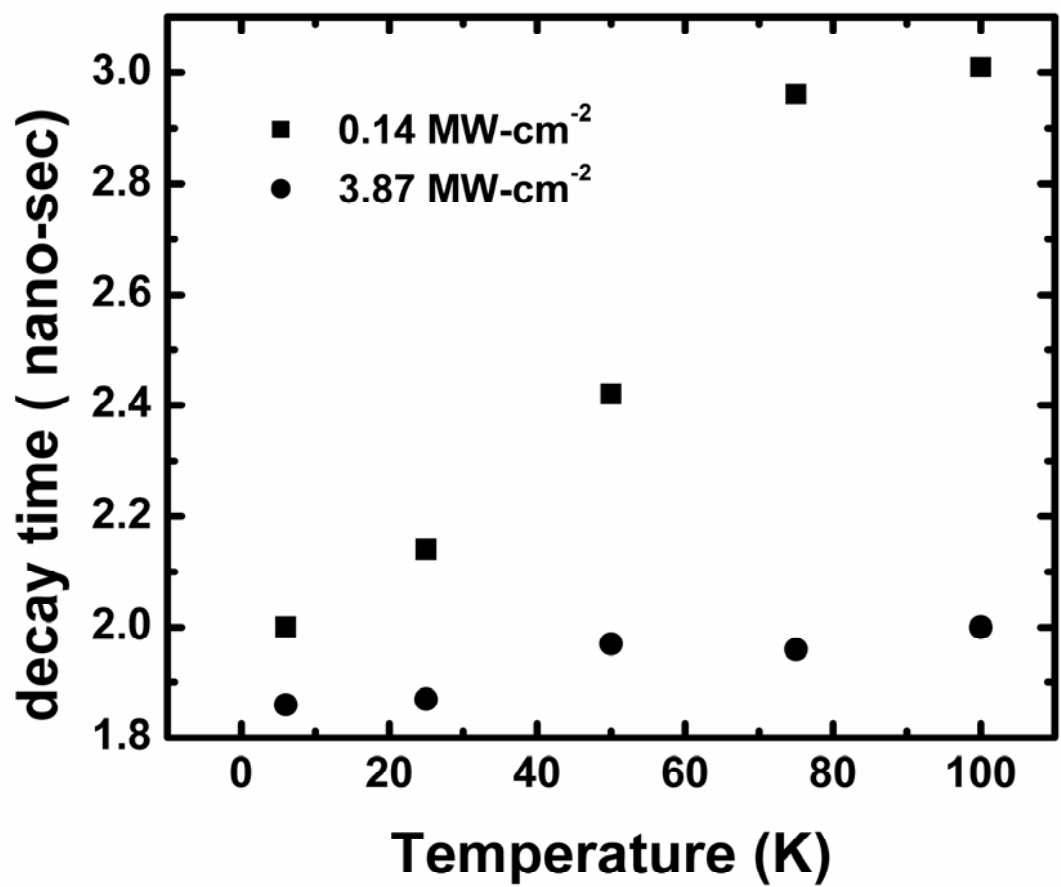

**Figure 5**

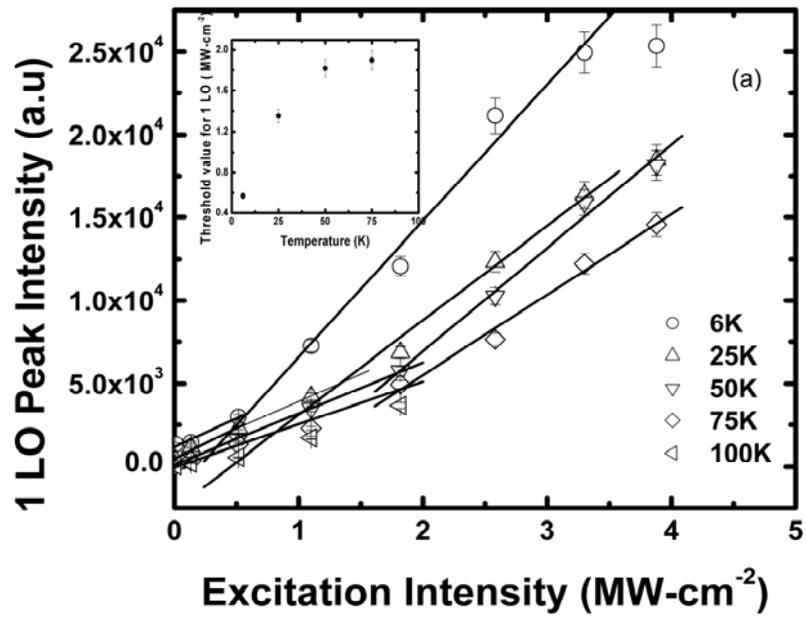

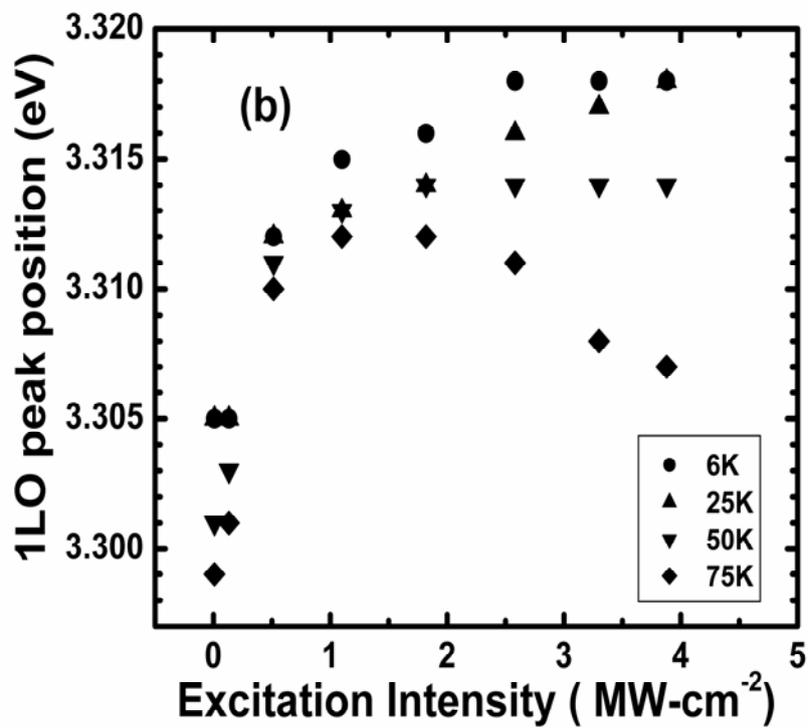

**Figure 6**